# A Feasibility Study on SNTP and SPoT Protocols on Time Synchronization in Internet of Things

Nelda Raju, Fida Hassan

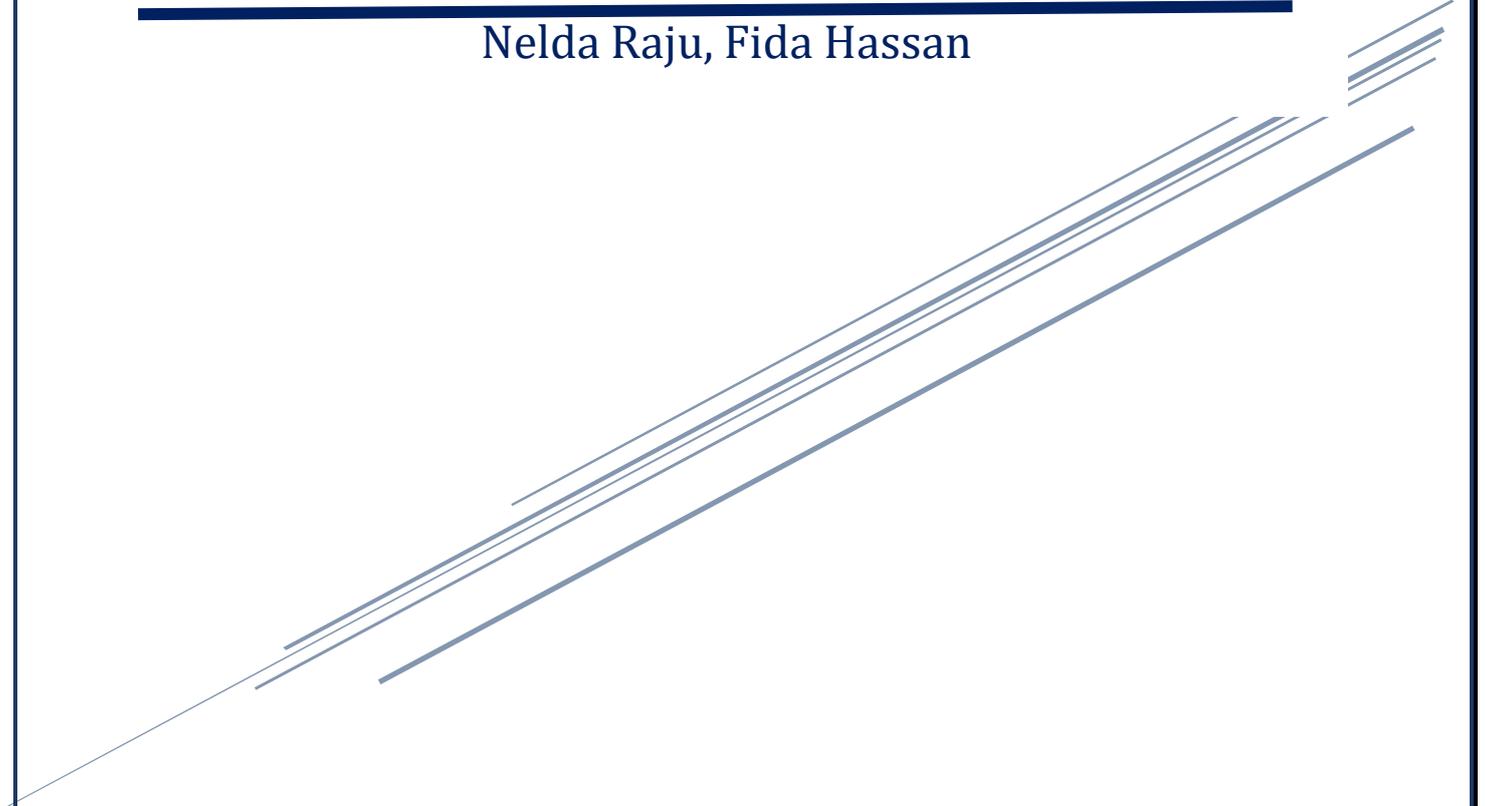

# Abstract

The new wave of computing allows users to explore their time in the Internet of Things (IoT) by connecting their smart devices over the network for data transfer without human interventions. While this swing increases the pace in IoT, time synchronization became a demanding feature on IoT devices for real-time applications. In this paper, we describe two synchronization protocols in IoT, its features and the advantages over the other. In addition, this work reveals the importance of time synchronization in the IoT platform and the effects of bad syncs in the real-time applications. We start our research by analysing the widely using Simple Network Time Protocol (SNTP) and its performance in terms of offset and delay to examine the accuracy and reliability to define the synchronization. We have thoroughly analysed the time calculations that happens during the synchronization and the different algorithms that involved in each protocol. We observe an offset of approximately 100 milliseconds from the UTC standards. In order to address the clock drift and the stability, we compared the performance parameters of SNTP with the newly developed synchronization system, Synchronization Protocol for ioT (SPoT). We evaluate the efficiency of both the protocols in different operating conditions including the variations in the noise level to find out the efficacy that performs in the IoT platform. Furthermore, our analysis leads us to grab our attention on the two core algorithms involved in SPoT synchronization system to maintain the high clock accuracy rate. We find that SPoT performs 17x more accurately than SNTP at various noise levels by maintaining a clock accuracy within approximately 15 milliseconds. Moreover, we have keenly observed the implementation methods of both SNTP and SPoT protocols and its requirements for the successful implementation. This paper details all the findings and contributions that we have derived throughout our research and the related works that define the future scope of SNTP and SPoT protocols in IoT platforms.

# Contents

# Introduction

## Background

Internet of Things (IoT) is a platform where billions of smart devices are connecting together over the network for data exchanges without human interactions. The self-operability of this invention is the main feature that attracts today's world in all aspects. A device that connected to the network itself will not be considered as an IoT device unless it transfers data without human interventions. An IoT device can be anything like a system enabled car, trolley, fridge, computer, smartphones and so on. This wide choice of applicability accelerates the growth of IoT platforms. In every year, billions of IoT devices invisibly connecting to the internet to communicate and transfer data. In short, this connects the digital and physical world through its digital intelligence.

Between these connections, synchronizations bring a lot of meaning for the real-time control applications like monitoring, security, tracking, and environmental sensing. Time synchronization is a process that adjusts the internal clocks of all the connected smart systems according to universal time standards. It is essential to have unique time settings for all the systems that are linking together over the network to achieve a specific goal.

This research reveals more about the time synchronizations in IoT and the two efficient protocols used for this to achieve a unique time throughout the network.

## Research Question

Internet of Things connects billions of IoT devices together for data sharing. In this scenario, this study opens a space to learn more about the time synchronization mechanism that implements on IoT platforms. IoT devices that connected over the network will be from different regions with varied time zones. Hence, to have an accurate time sync between all the devices is really essential for most of the IoT applications especially the applications that deal with real-time data sources. Therefore, a study on this need is a relevant attempt. Time synchronization protocols are the mechanisms that help the IoT platform to address these time sync issues. The commonly used protocols for IoT platforms are, Network Time Protocol (NTP) for routers, Precision Time Protocol (PTP) for computer, Simple Network Time Protocol (SNTP) for smartphones and computers, Datacentre Time Protocol (DTP) for data centres and Synchronization

Protocol for IoT (SPoT). The researchers claim that the newly developed SPoT protocol performs more accurately than SNTP due to its effective synchronization algorithms and filtering capabilities. Since the SNTP is one of widely used protocol for the enterprise applications, it is significant to have a detailed study on the above statement. By considering the importance of this topic, my research question deals with the below contexts.

- How the SNTP and SPoT protocols maintain the time synchronization in IoT?
- What kind of time calculations are behind these synchronization protocols?
- Why the researchers prefer SPoT over the SNTP?
- What are the parameters that determine the good sync in SNTP and SPoT protocols?

This work is a research attempt to ensure the best potential solution for the above queries.

## Aims and Objectives

The overall purpose of this study is to have a feasibility study on SNTP and SPoT protocols on time synchronization in the IoT platform. The complete work has been carried out in three distinct processes. The first stage was all about the literature review on IoT and Time Synchronization. The next stage is an analysis of various time sync protocols to define two widely used protocols for the feasibility study. This stage results in the selection of SNTP and SPoT protocols for the further research. The final stage is to perform the actual research to ensure the evaluation of SNTP and SPoT protocols and its future scopes. The anticipated outcome of this project is to define the parameters that describe the statement SPoT performs more accurately than SNTP.

Our goals include,

- Review on the existing literatures to acquire the in-depth knowledge on time synchronization protocols and IoT platforms
- Research to define the parameters that determine the efficiency of SNTP and SPoT protocols.

IoT is an outcome of the digital revolution. The main significance of this study itself is the wide applicability of this platform. The objective of this study is to define the best available time synchronization protocol in IoT platform.

### Overview of the method

This study is a research attempt to define the parameters that determine the accuracy between the time synchronization protocols SNTP and SPoT. The recent research claims that SPoT performs 17x more accurately than SNTP protocol. This attempt is exactly to identify the clock accuracy of SNTP when compared with SPoT and to ensure the above statement remains true or not. For that, we are planning to have a practical implementation of SNTP over the network and evaluate its performance in terms of clock offset and delay. Clock offset is the parameter which defines the difference in time between two machines. This value defines how accurately the systems are coupled together and transfers data with each other. Due to the laboratory restrictions, we cannot have an implemental research of SPoT synchronization system, whereas will compare the values of SNTP with the research values of SPoT to retrieve the findings and observations.

### Scope of the project

The section background, research question and aims and objectives collectively specifies the significance of this study in the IoT platform. The current generation is behind the digital world, where they cannot exist without the internet. Nowadays IoT devices became a communicating node which serves numerous information through various real-time applications. The scenario is like everything in this world is just around your fingertip. One of the world's leading research and advisory company, Gartner, states that, in 2017, there are around 8.4 billion IoT devices were connected across the world, which is 31% more in the count than 2016. He also mentions that this count will reach up to 20.4 billion by 2020 (Gartner, 2017). These striking figures show the need for this kind of studies at this point. Time synchronization is really important to save our battery life, to have a secured connection, for immediate results and for real-time applications (Burns, 2017). This project scope is very specific on time synchronization in IoT and very useful for the growing IoT platform.

### Key deliverables

The ultimate outcome of this study will be a research finding which defines the statement SPoT protocol performs 17x more accurately than SNTP. The deliverable includes the observations from both literature review and the experimental research. The results point out the potential solution for the research question raised at the beginning of the project like the time calculation behind both SNTP and SPoT, and the parameters that define the accuracy of the synchronization protocols. In addition, the research delivers the evidences for this feasibility study results.

# Literature Review

Internet of Things connects the physical world to the digital world by transferring data between them. In every year the width of this platform is increasing due to the increased count of networked devices over the network. The devices with various time zones are getting connected to the existing networks. Hence, synchronization is very important to keep the network within unique time settings. If not, it will affect the system in many ways. Synchronization is a process of adjusting the internal clocks of the networked devices according to the timestamps of the reference server. The main five reasons behind the need for time synchronization are security, battery life, real-time applications, need of immediate results and devices with different time zones (Burns, 2017). The solution for this issues is in the form of various network protocols available for IoT devices. Network Time Protocol (NTP), Simple Network Time Protocol (SNTP), Precision Time Protocol (PTP), Datacentre Time Protocol (DTP), Mobile Network Time Protocol (MNTP) and Synchronization Protocol for ioT (SPoT) are few of those synchronization protocols (Sathya Kumaran Mani, 2018). Out of these, routers and wired hosts in the network use the NTP protocol to sync the time with the reference clock. SNTP and MNTP are the protocols that are widely used by mobile phones and computers in the network. DTP is exclusively for data centres of the network. And the SPoT is for all the IoT devices that have been connected over the network. Out of these different protocols, we have chosen the SNTP and SPoT protocols for the research analysis.

The lack of synchronization can be overcome by adjusting the clock drifts and offset between the nodes in the network with respect to a universal standard time (Khondokar Fida Hasan, 2018). This helps the node operations to carry out in the same notion of time. Clock drifts are the differences in the clock rates of a practical clock when compared to the ideal clock. Below figure clearly shows the drift variations in the practical clock.

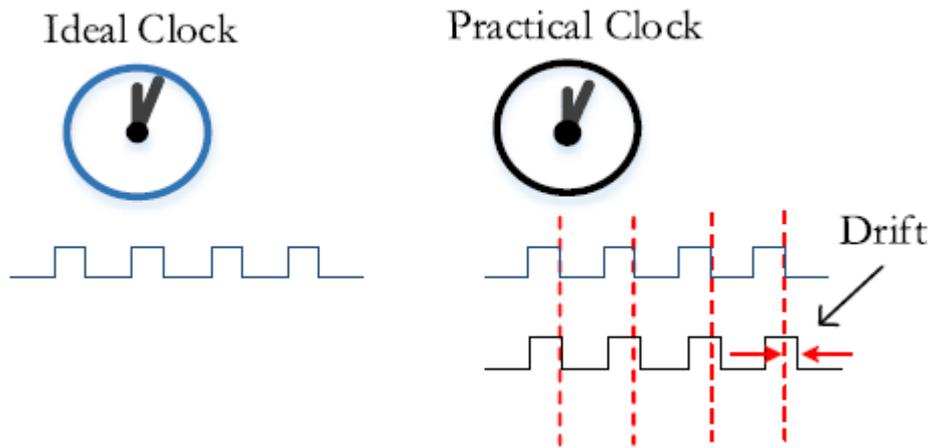

These small variations will make a difference in the node timings over the network. This difference is known as offset. The above figure shows this variation. The second differences in each nodes questions the efficiency of the system because the time sync is really necessary for the applications in the IoT platforms.

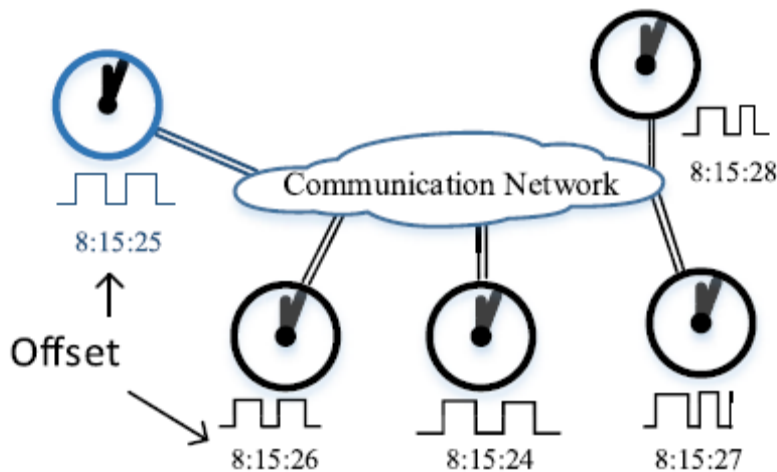

The communication between the nodes will help the system to reduce the clock drifts and offset errors in the network. This helps to transfer the node time from one node to another and update the nodes as per the reference time logged into the message packet.

From the initial analysis, we analysed that Simple Network Time Protocol (SNTP) is a client version of Network Time Protocol (NTP), where it does not provide the NTP feature like complex filtering and statistical mechanism (Cisco.com, 2018). Even though SNTP is a simplified version of NTP, we cannot configure both in the same system since they use the same port. Another key feature of the SNTP is that it evaluates the clock accuracy on the symmetrical scenarios and does not consider the asymmetrical traffic situations. When compared the SNTP with SPoT protocol, SPoT considers both symmetrical and asymmetrical scenarios over the network. The supreme feature of this protocol is to address the clock drift and clock rate stability. It is a lightweight protocol with high computational capabilities and accuracy parameters. Spot enables the filtering mechanism over the network to have an effective time synchronization. Let's have a look on the main parameters that involved in the time synchronization in both SNTP and SPoT protocols.

**Time Calculation**

The time calculations behind the SPoT protocol is the same as in the SNTP. In these protocols, time synchronization is implementing through a set of packet exchanges. In every packet exchange, four variations of timestamps will be recorded in the packet. They are Originate Timestamp, Receive Timestamp, Transmit Timestamp and Receipt Timestamp. The below figure details about this timestamps in terms of a client request and server response exchange.

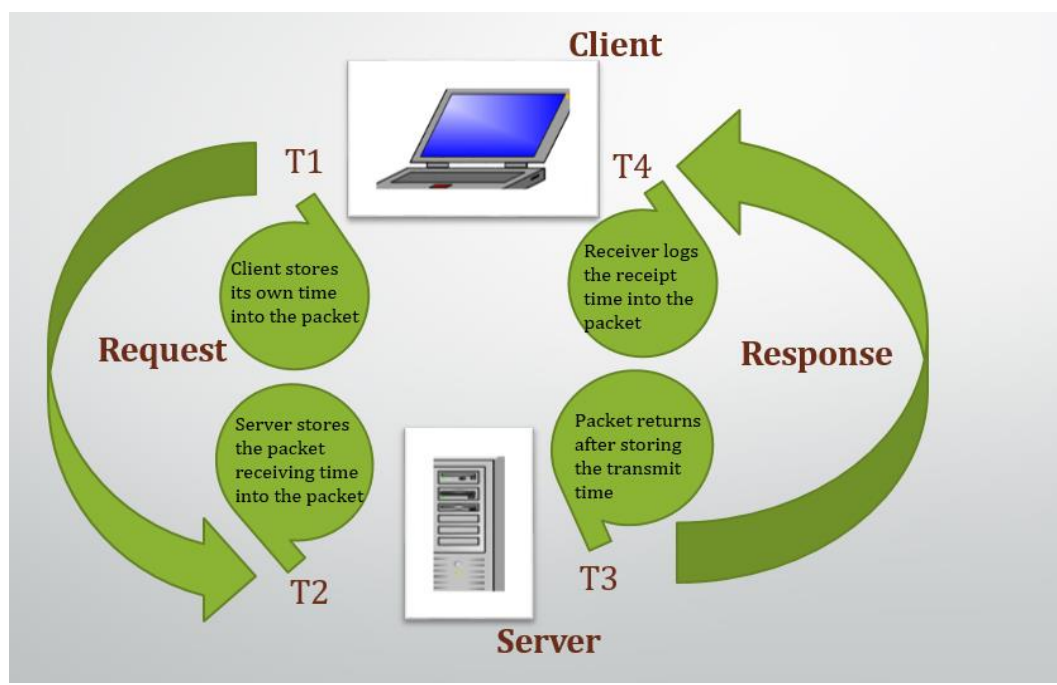

- Originate Timestamp (T1) – The timestamp that has been recorded into the packet by the client when the client initiates a request to the server.

9 | P a g e

- Receive Timestamp (T2) - The timestamp that has been logged into the packet by the server when the server receives the request from the client.
- Transmit Timestamp (T3) - The timestamp that has been recorded into the packet by the server when the server sends back the response to the client.
- Receipt Timestamp (T4) - The timestamp that has been logged into the packet by the client when the client receives the response from the server.

The total delay of the packet exchange can be calculated by subtracting the originate timestamp from the receipt timestamp (T4 – T1) and the remote processing time can be calculated by subtracting the receive timestamp from the transmit timestamp (T3 – T2). From all these calculations, we can derive the travelling time, roundtrip delay and clock offset as follows.

$$\text{Travelling time} = \tfrac{1}{2}(\text{Total delay} - \text{Remote Processing Time})$$
$$= \tfrac{1}{2}((T4-T1) - (T3 - T2))$$

Let a = T2 - T1 and b = T3 – T4
- Roundtrip Delay = a-b
- Clock Offset = (a+b)/2

The above calculations take the scenario in a symmetrical exchange. But in reality, there could be chances to occur an asymmetrical scenario due to the switching delays or dynamism. Since most of the IoT devices performed its activities through a wireless mode, this change is more often to occur. When an asymmetrical delay occurs, SPoT protocol collects multiple reference time to figure out the differences.

### Clock Offset

Clock offset is the value that describes the quality of time in the network. By evaluating the clock offset, we can define and compare the accuracy of each protocol in an IoT platform. These drifts and offset values will vary in different devices based on the environmental conditions. The clock drift values have a great impact on the room temperatures where the IoT device is been placed. Recently a research has been conducted to identify this influence in the IoT devices. The results noted that the different hardware instances shows different drift rate in the same platform and also the same hardware show different behaviour with respect to variant room temperatures (Sathya Kumaran Mani, 2018). Below figure shows the specified findings in details.

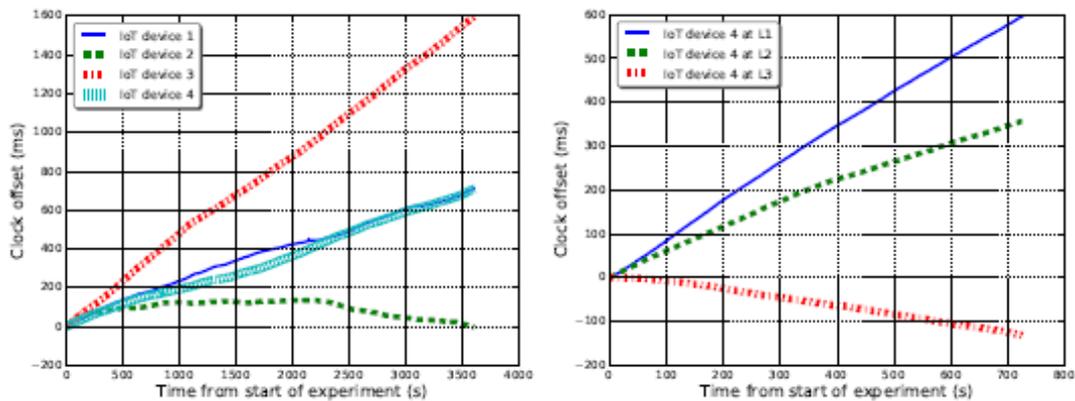

The first figure indicates the offset values of different IoT devices in the same location. We can see the huge differences between device two and three. Whereas the second figure is the offset values of same device one, two and three at three different locations with different room temperatures. Device one is at the server room temperature of 14 degree Celsius, device two is at 21 degree Celsius and device three is at 27.5 degree Celsius. The device two and device three locations have been selected to represent the office room and residential apartment respectively. These locations represent the real IoT scenario and the results were shocking. The huge variations were shown by the devices with inexpensive hardware clocks.

**Algorithms Involved**

SNTP protocol implementation and processing are based on the four main algorithms. The offset analysis between nodes is really important for the good time synchronization protocol. For considering accurate and most reliable subsets of these offset values or delay values, SNTP follows the peer selection algorithm. All the time calculations in SNTP will be carried out based on the Data-Filtration algorithm that has been implemented in it. DES encryption algorithm will help the SNTP enabled the network to have a proper authentication between every node to prevent the unauthorized access from the external sources. A proper routing protocol is essential to have reliable and faster packet exchanges. Bellman-Ford algorithm will help the network to choose the most reliable and shortest set of nodes for the required data transfers in an efficient way. SPoT protocol, mainly have two algorithms named offset synchronization algorithm and rate synchronization algorithm. The main idea behind the offset synchronization in SPoT is to address the asymmetry errors. In order to address this error, we need to identify the directions of asymmetry and the magnitude of the same. The asymmetry can happen either in the forward direction or in the backward direction. Once the direction identifies the algorithm corrects the difference in the offset calculation to have the proper time updates. The delay that happens in the client to server exchange is termed as forward asymmetry and in the server to client exchange is termed as the backward asymmetry. The direction identification is based on the offset variations. The increase in the expected offset indicates the forward asymmetry and the decrease in the same notifies the backward asymmetry. Hence by validating the computed and expected offsets results the direction of asymmetry error in the network. The magnitude of this error will depend on the difference between the minimum round trip delay that has already logged in the system and the current sample.

```
input: measuredOffset = offset from measurement
input: measuredRTT = RTT of measurement
input: oldOffset = last known offset
input: timeDelta = duration since last measurement
input: clockSkew = estimated rate-error of clock
input: minRTT = minimum RTT seen
input: errorMargin = tunable error margin
Function filterOffset()
    estimatedOffset = oldOffset + clockSkew * timeDelta
    asymmetricDelay = measuredRTT - minRTT
    if measuredOffset > estimatedOffset + errorMargin then
        // forward asymmetric error
        correctedOffset = measuredOffset - 0.5 * asymmetricDelay
    else if measuredOffset < estimatedOffset - errorMargin then
        // reverse asymmetric error
        correctedOffset = measuredOffset + 0.5 * asymmetricDelay
    else
        // symmetric additional delay
        correctedOffset = measuredOffset
    return correctedOffset
```

As per the algorithm represents, is the measured offset is greater than the summation of the estimated offset and error margin, then there is a chance of forward asymmetric error. So the algorithm needs to fix the offset by subtracting the half of asymmetric delay. Whereas if the measured offset is lower than the estimated offset then this is the case of a backward asymmetry and the algorithm will correct the offset by adding the half of asymmetric delay along with the measured offset. If the above two conditions fail, then it will be a symmetric scenario where the corrected offset remains the same as the measured offset. Now the corrected offset is the parameter which holds the value that determines the quality of the network time. Another important algorithm in SPoT protocol is the rate synchronization algorithm. The difference between two clock rates has been termed as clock skew. This is one of the values that calculates the estimated clock offset and correcting the clock drift. The rate synchronization algorithm in SPoT will help to perform these in the network with its features.

```
input: measuredOffset = offset from measurement
input: errorMargin = tunable error margin
input: pollingStyle = AIMD or MIMD as chosen
Function synchronizeClockRate()
    if In observation time or numSamples <5 then
        absoluteError = abs(estimatedOffset - correctedOffset)
        meanAbsoluteError.update(absoluteError)
        numSamples += 1
    else if meanAbsoluteError <2 * errorMargin then
        // clock has been stable so far
        // increase polling interval
        pollingInterval.increase(pollingStyle)
        restartObservationTime()
        numSamples = 0
    else
        // clock is unstable
        // decrease polling interval
        pollingInterval.decrease(pollingStyle)
        restartObservationTime()
        numSamples = 0
    if correctedOffset = measuredOffset then
        // high quality offset sample
        // update clock skew
        clockSkew = (clockSyncOffset - correctedOffset) / (clockSyncTime - currentTime())
        clockSyncOffset = correctedOffset
        clockSyncTime = currentTime()
```

The algorithm first checks the stability of the clocks. If it finds the clock has been stable then it increases the polling interval and if it finds the clock has been unstable and decreases the polling interval. The point where the corrected offset and measured offset become equal, there we can see the high quality offset samples and can update the clock skew. Whenever the offset algorithm executed, the rate synchronization algorithm also executes simultaneously.

# Project Methodology

Our research is a comparative study of SNTP and SPoT protocols, we have divided the experimental study into two parts. The first part includes the implementation of the SNTP protocol and analyses the values that determine the time accuracy in it. This part is purely an experimental approach. Whereas the second part is the idea that deals with the SPoT protocol. Due to the laboratory restrictions, we have chosen the SPoT implementation analysis part from the existng research paper.

## SNTP Implementation and Analysis

The first step for this was to set up the SNTP server. I have tried this setup in both windows and Linux operating systems to have a better clarity. First will have a look on the SNTP server setup in the windows machine.

<u>In Windows</u>

At first, we need to initiate the windows time in the windows machine. For that, we have to follow the below steps.

Start –> Select Control Panel –> Select Administrative Tools options –> Select Services –> Select Windows Time from the list–> Right Click on Windows time -> Click Start

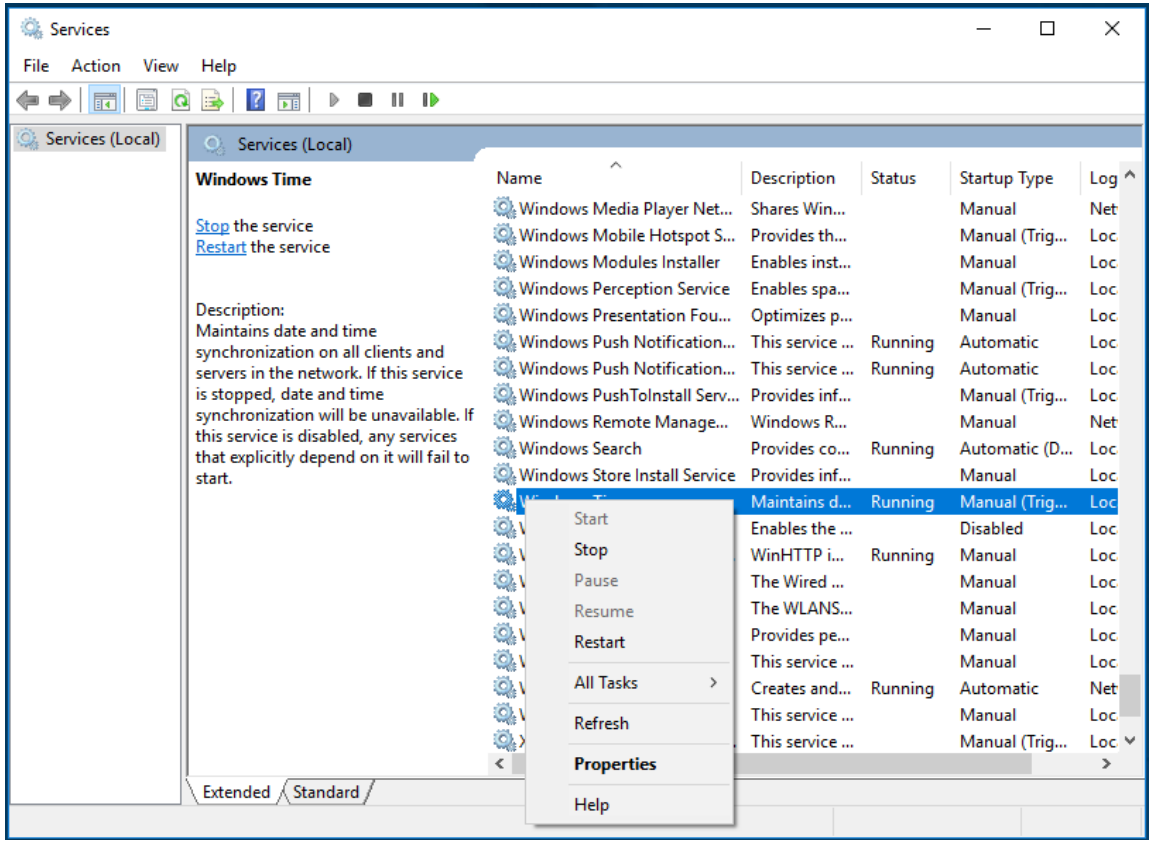

Now the windows time is up and running. Next we need to edit the registry properties in Regedit. In Regedit,

Select HKEY local machine –> Select System –> Select Currentcontrolset – >Select Services – >Select W32 Time –> Select Time Providers –>Select NTP server –> Double click on Enable option and change the value to 1 –> Click Ok

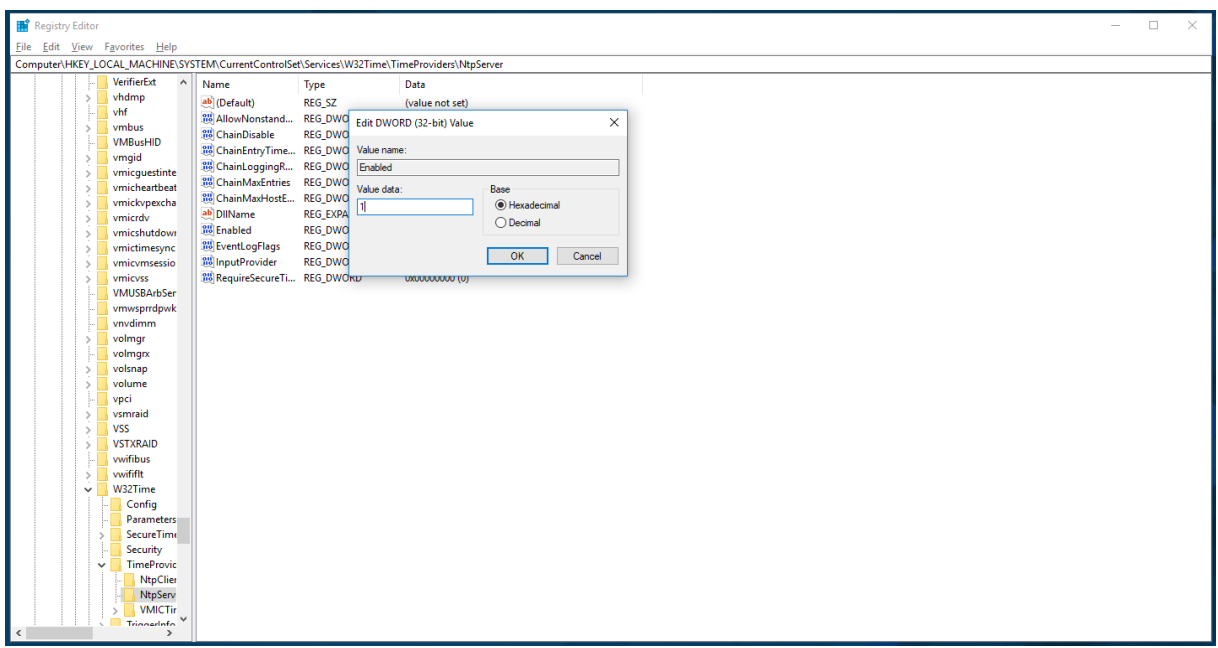

The next step is to set the flag values. For that in W32time,

Click on Config Folder –> Double click on Announce Flags –> Change the value to 5 –> Click Ok –> Close the registry editor

Once till the flag settings finished, we need to unblock two ports. They are UDP incoming port and UDP outgoing port. Since the SNTP uses UDP for packet exchange in the transport layer and the port as 123 we need to follow the below settings to set up the windows machine as SNTP server.

Unblock port 1:

Start –> Select Control Panel –> System and security –>Windows firewall –> Advanced settings –> Inbound rules –> New rule –> Port –> Click Next –> Select UDP –> Type 123 -> Click Next –> Allow the connection –> Click Next –> Name as udp123incoming –> Click Finish

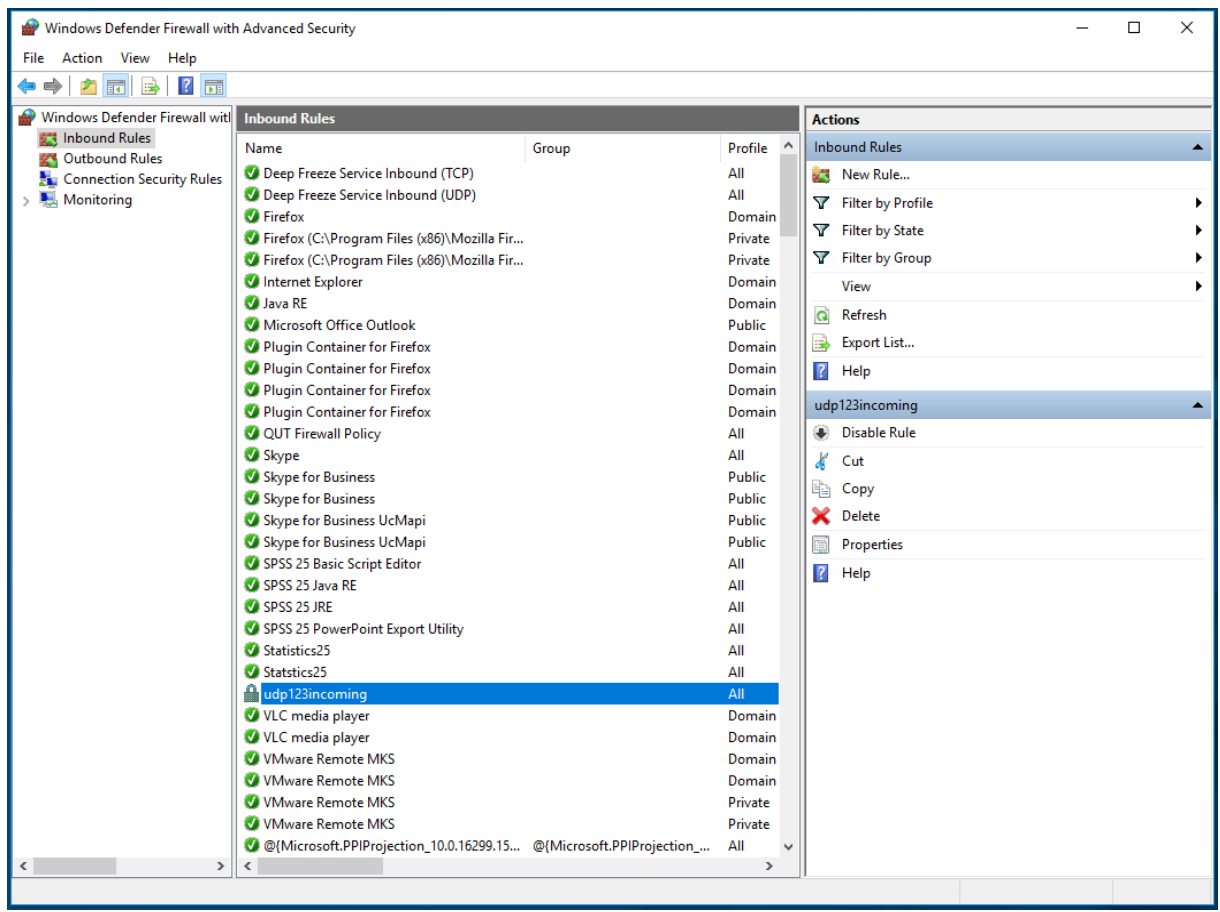

Unblock port 2:

Start –> Select Control Panel –> System and security –>Windows firewall –> Advanced settings –> Outbound rules –> New rule –> Port –> Click Next –> Select UDP –> Type 123 –> Click Next –> Allow the connection –> Click Next –> Name as udpoutbound123 –> Click finish –> Close the firewall window

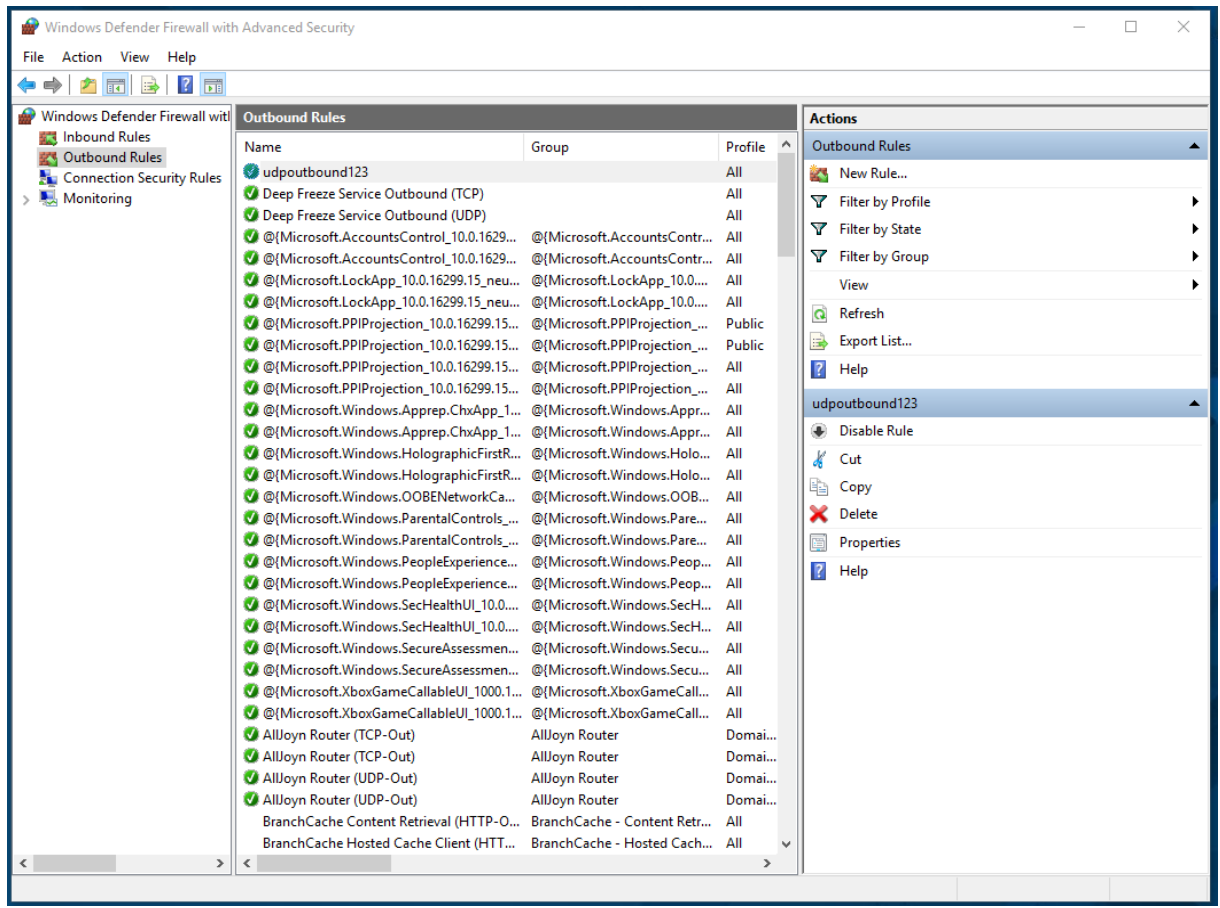

After unblocking the two ports, we need to start windows time again by

Start –> Control panel –> System and security –> Administrative tools – > Services –> Scroll down and find Windows Time –> Right click on windows time and click start

Once it is done, we need to modify the windows time by

Right click on Windows time –> Click on properties –> Change the status type as automatic –> Apply –> Click Ok

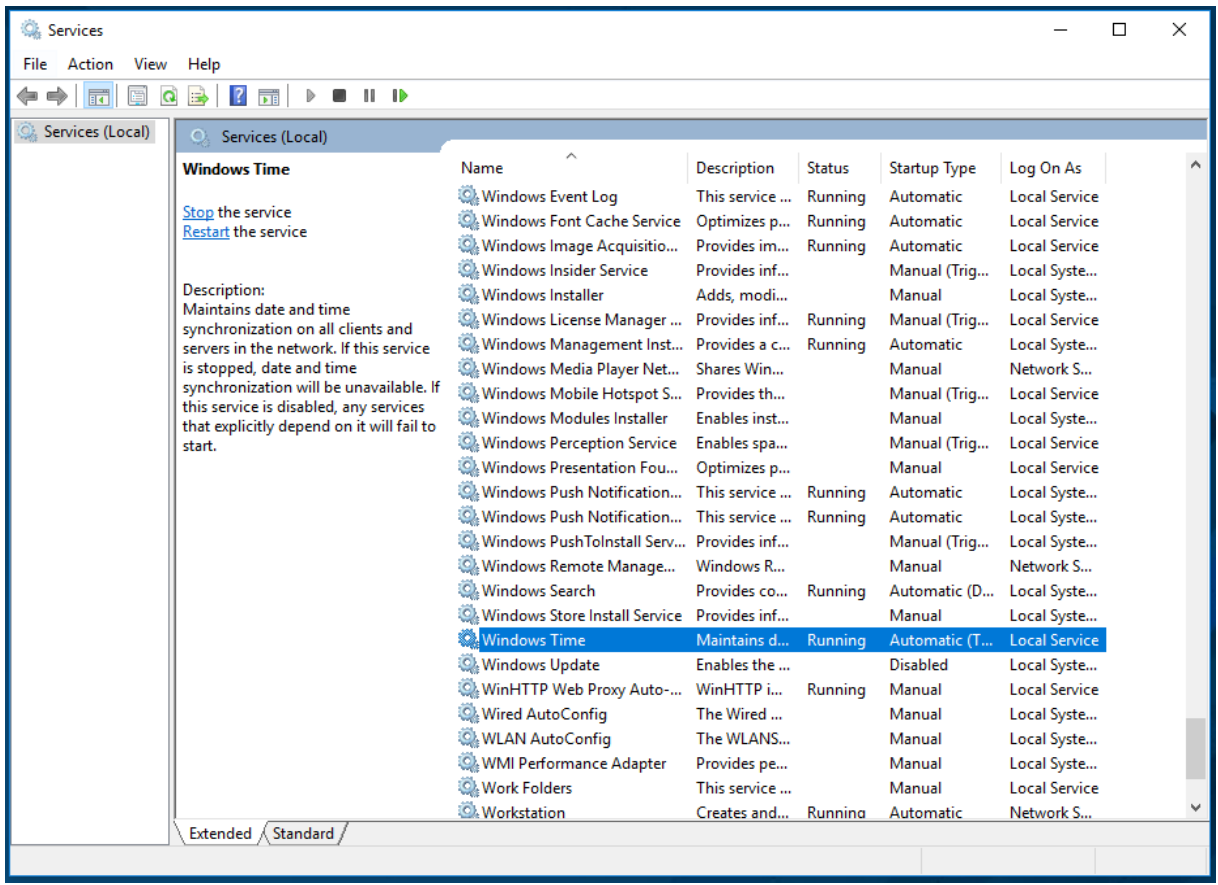

Now the SNTP server set up has been done in windows and we can evaluate the time in the server using the ntpserver tool or through the command prompt, we can view the SNTP synchronised time and stratum server details. W32tm /query/status provide us with the below result which specified the stratum server as 6 and synced by SNTP.

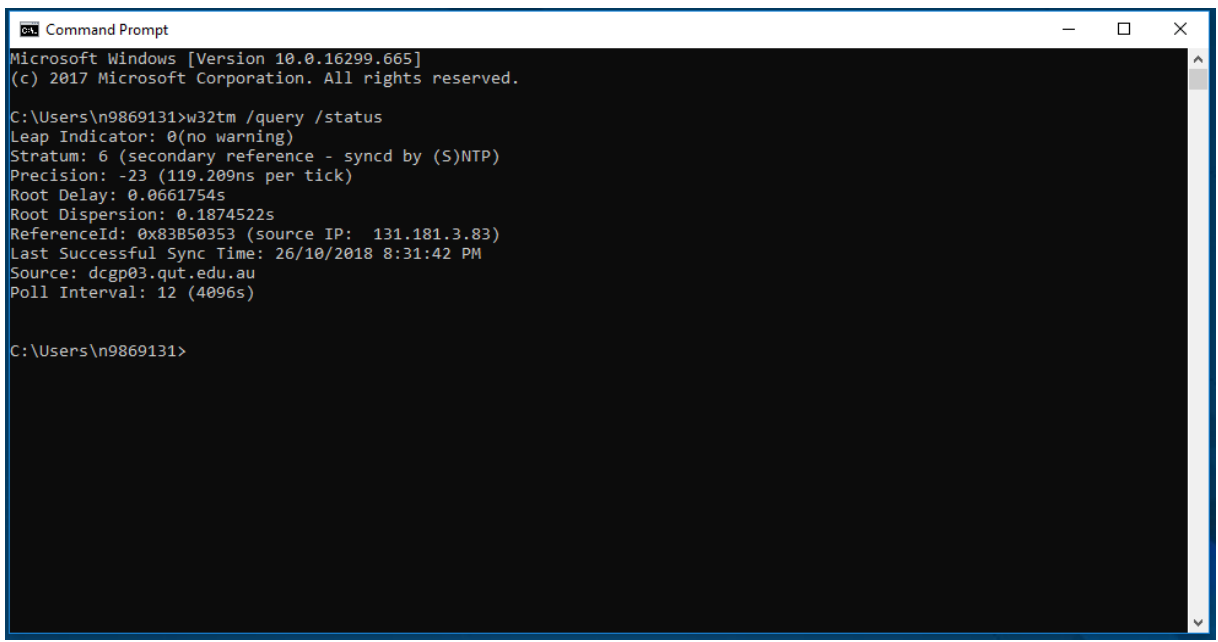

In Linux

With the help of apt-get update and install features, we can easily download the SNTP protocol framework in our Linux system. Once the download completes go to the nano editor and edit the configuration file to configure the server.

```
driftfile /var/lib/ntp/ntp.drift

# Enable this if you want statistics to be logged.
#statsdir /var/log/ntpstats/

statistics loopstats peerstats clockstats
filegen loopstats file loopstats type day enable
filegen peerstats file peerstats type day enable
filegen clockstats file clockstats type day enable

# Specify one or more NTP servers.

# Use servers from the NTP Pool Project. Approved by Ubuntu Technical Board
# on 2011-02-08 (LP: #104525). See http://www.pool.ntp.org/join.html for
# more information.
#pool 0.ubuntu.pool.ntp.org iburst
#pool 1.ubuntu.pool.ntp.org iburst
#pool 2.ubuntu.pool.ntp.org iburst
#pool 3.ubuntu.pool.ntp.org iburst

# Use Ubuntu's ntp server as a fallback.
#pool ntp.ubuntu.com
#pool time.admin.ifn641
server 0.ubuntu.pool.ntp.org iburst
server 1.ubuntu.pool.ntp.org iburst
server 2.ubuntu.pool.ntp.org iburst
server 3.uduntu.pool.ntp.org iburst
server ntp.ubuntu.com iburst

# Access control configuration; see /usr/share/doc/ntp-doc/html/accopt.html for

webserver@webserver:~$ systemctl reststart ntp
Unknown operation reststart.
webserver@webserver:~$ sudo systemctl restart ntp
webserver@webserver:~$
```

Start the NTP daemon to verify the synchronization is active or not. The below screen showing the status as active and running.

```
webserver@webserver:~$
webserver@webserver:~$
webserver@webserver:~$ ip netns exec vrouter ntpq -c lpeer 127.0.0.1
Cannot open network namespace "vrouter": No such file or directory
webserver@webserver:~$ ip netns exec ntpq -c lpeer 127.0.0.1
Cannot open network namespace "ntpq": No such file or directory
webserver@webserver:~$ sudo systemctl restart ntp
webserver@webserver:~$ sudo systemctl status ntp
● ntp.service - LSB: Start NTP daemon
   Loaded: loaded (/etc/init.d/ntp; bad; vendor preset: enabled)
   Active: active (running) since Tue 2018-10-16 19:37:32 AEDT; 10s ago
     Docs: man:systemd-sysv-generator(8)
  Process: 25185 ExecStop=/etc/init.d/ntp stop (code=exited, status=0/SUCCESS)
  Process: 25199 ExecStart=/etc/init.d/ntp start (code=exited, status=0/SUCCESS)
    Tasks: 2
   Memory: 628.0K
      CPU: 15ms
   CGroup: /system.slice/ntp.service
           └─25212 /usr/sbin/ntpd -p /var/run/ntpd.pid -g -u 111:119

Oct 16 19:37:32 webserver ntpd[25212]: proto: precision = 0.061 usec (-24)
Oct 16 19:37:32 webserver ntpd[25212]: Listen and drop on 0 v6wildcard [::]:123
Oct 16 19:37:32 webserver ntpd[25212]: Listen and drop on 1 v4wildcard 0.0.0.0:123
Oct 16 19:37:32 webserver ntpd[25212]: Listen normally on 2 lo 127.0.0.1:123
Oct 16 19:37:32 webserver ntpd[25212]: Listen normally on 3 ens160 10.11.1.14:123
Oct 16 19:37:32 webserver ntpd[25212]: Listen normally on 4 ens160 10.11.1.5:123
Oct 16 19:37:32 webserver ntpd[25212]: Listen normally on 5 lo [::1]:123
Oct 16 19:37:32 webserver ntpd[25212]: Listen normally on 6 ens160 [2402:ec00:ffed:1:5::]:123
Oct 16 19:37:32 webserver ntpd[25212]: Listen normally on 7 ens160 [fe80::250:56ff:feae:787c%2]:123
Oct 16 19:37:32 webserver ntpd[25212]: Listening on routing socket on fd #24 for interface updates
webserver@webserver:~$ sudo ntpq -np
```

Once the SNTP service is up and running, we can retrieve the offset and delay values through ntpq query as shown below

```
webserver@webserver:~$ sudo ntpq -c lpeer
     remote           refid      st t when poll reach   delay   offset  jitter
==============================================================================
 0.ubuntu.pool.n .POOL.          16 p    -   64    0    0.000    0.000   0.000
 1.ubuntu.pool.n .POOL.          16 p    -   64    0    0.000    0.000   0.000
 2.ubuntu.pool.n .POOL.          16 p    -   64    0    0.000    0.000   0.000
 3.ubuntu.pool.n .POOL.          16 p    -   64    0    0.000    0.000   0.000
 ntp.ubuntu.com  .POOL.          16 p    -   64    0    0.000    0.000   0.000
webserver@webserver:~$ clear
```

For this study, I have continuously retrieved the delay and offset values to observer the clock accuracy and quality of time.

### SPoT Implementation and Analysis

SPoT protocol supports both thick and thin clients. The thick client is the application that has to be installed in the local system and it uses the computer resources for the successful execution. Whereas the thin client is the web application that we can access through the browser and it uses the internet. In short, the thick client is a client-side application and the thin client is a server-side application (Webopedia, 2006). The synchronization algorithm of SPoT protocol will be running in the client side for the thick client and for the thin client the algorithms will execute on the server side only. To perform these two functionalities, SPoT has four key components including Core library for thick clients, Scalable reference server for both thick and thin clients, the reference implementation for thin client and a client emulator. This protocol supports both C and C++ languages since the core library has around 400 lines of C code for the implementation and the reference server and client emulator has around 300 and 130 C++ code line respectively. For the spot evaluation, the researchers have taken the experiment at different noise levels to analyse the effects on offset at various noise levels. For this analysis, we are not considering the network errors like packet drop and duplications since it will not affect the clock accuracy in any manner. Here, we are strictly concentrated on the low, medium high noise levels and the preset noise distribution deviations. The standard deviation for this has been set as 50 milliseconds for low noise level, 150 milliseconds for medium noise levels and 250 milliseconds for the high noise levels. We are not comparing the thick and thin client performance for this study since they both perform the same in terms of accuracy.

The Spot architecture is very simple to understand. The IoT devices that wish to adopt the SPoT facility needs to register with the SPoT server well in advance along with their polling style, device type and EM. Since the Thin client can access through the browser, the server will respond to the client request along with the timestamp. For this, the algorithm will execute in the SPoT server. For the thick client, it has to execute the algorithm in the local machine where it got installed and required to get the timestamp from the SPoT server through packet exchange. The two main algorithms and 4 major parts are the component that included in the SPoT architecture.

# Outcomes

Internet of Things is a platform with a large number of devices connected over the network. The main idea behind this system is that there will not be any kind of human interventions. This study enables me to earn a strong base on IoT concepts and time synchronization mechanisms. Moreover, I got introduced with different time synchronization protocols and their systems. Out of these many protocols, I have chosen the SNTP protocol for the analysis due to tie wide acceptance and SPoT protocol due to its proved performance. After gaining the base on these two protocols, there were few queries arises, I reframed it into 4 research questions which address the time calculations in IoT, algorithms used and parameters for the good time synchronization in IoT based on the SNTP and SPoT protocols.

At the beginning of this research study, I have addressed four research questions that analyses the SNTP and SPoT protocols. The first and foremost questions was about the time synchronization of both protocols in IoT platform. The main observation from this research for this query is that for SNTP protocol, it follows the reference server or stratum server concept for the time synchronization. That means, the servers in the platform will be divided into different stratum levels and name the levels from stratum 0, stratum 1, to stratum n servers. The stratum 0 servers are the reference servers with the UTC time zones. All other secondary stratum servers updates their internal clock with respect to the stratum 0 servers. This update or synchronization is happening through the packet exchange between the nodes. When a request and reply exchange happen between nodes in the network the timestamp will be logged in the packet. After calculating the RTT and offsets the secondary servers will update their clock. When compared this mechanism with the SPoT protocol, we can see a similar approach in the synchronization method. But there is no stratum server concept in SPoT architecture. Instead, the system will be having a SPoT server which possess a global time zone. The device or client which wish to be part of it needs to register to the SPoT server by providing few device details such as device type, polling style and EM. Since the SPoT architecture accepts both thick and thin clients the synchronization method is different for both clients. For thin clients, after the registration SPoT server initiates a timestamp exchange. From the client part, after considering the offset and clock skew, it will adjust their internal clocks accordingly. All the synchronization algorithms will be executed in the server side for the thin clients since it can be access through the browser over the internet. Whereas for the thick clients, after registration, it needs to execute the synchronization algorithms and need to share its timestamped packet to the SPoT server based on the algorithms. SPoT server in turn respond to the client with the global timestamp and then client will adjust their clocks accordingly.

The second question for this study was about the time calculations behind these protocol for achieving the clock accuracy. My research provides me with a clear outcome for this question. The time calculations are mainly based on the algorithms used in each protocol. But the common parameters for these algorithms will be calculated based on the four main timestamps that have been recorded during the packet exchange between client and server. The originate timestamp, Receive timestamp, Transmit timestamp and receipt timestamp will determine the rest all calculation for the algorithms. These calculations are nearly the same for both SNTP and SPoT protocols. With the help of above-mentioned timestamps, the framework will be able to calculate the travelling time. That is the half of the value that we get after subtracting the remote processing time from the total delay. Round trip delay and the clock offset are the other two parameters that influence the time synchronization algorithms. Let's name the timestamps as T1 for originate timestamp, T2 for receive timestamp, T3 for Transmit timestamp and T4 for receipt timestamp. The RTT and offset can be calculated as shown below

Let a = T2 - T1 and b = T3 – T4
- Roundtrip Delay = a-b
- Clock Offset = (a+b)/2

The only difference that has been observed during the research is that SNTP calculations are based on the symmetrical scenarios whereas SPoT protocol considers both symmetric and asymmetric errors. For evaluating both the scenarios, SPoT offset synchronization algorithm compare the measured offset and estimated offset. If the measured offset is greater than the estimated offset then the algorithm identifies the forward asymmetric errors and if the measured offset is lower than the estimated offset then the algorithm identifies the backward asymmetry. In order to correct the offset for the forward asymmetry, algorithm deducts the half of the asymmetry delay from the measured offset. And for the backward asymmetry, algorithm add the half of the asymmetry delay to the measured offset.

The third question in my research is about the preference of SPoT over the SNTP protocol. As we all know that SNTP is the most commonly used network time protocol in the IoT platforms. When compare the SPoT with SNTP, Spot is the newly developed protocol that is implemented into the field to conquer the IoT platform with its unique time synchronization features. The major advantage of SPoT over SNTP is that it suits for all types of IoT devices that vary in terms of its budgets, requirements, computational capabilities and so on. The algorithms that supported by the SPoT protocol are light weighted and supports a wide range of devices to synchronize their clock to the SPoT server. The flexibility and extensibility are another important factors that makes SPoT a bit more preferred than the SNTP. Beyond all these advantages, the simple logic behind the synchronization and the architecture makes SPoT to a new level in the IoT platform. The factors that mentioned above is all the advantages of SPoT over SNTP. But the real component that makes SPoT to conquer is that the clock accuracy within 10 milliseconds. SNTP is having around 100 milliseconds of clock accuracy, which is really lower than the accuracy rate shown by the SPoT protocol. The offset value is the main parameter that determines the quality of time in networking protocols. By analysing these values and deviation in these values, we can compare the performances of the protocols. As per the research work that we have conducted, we have analysed the offset values of both SNTP and SPoT protocols.

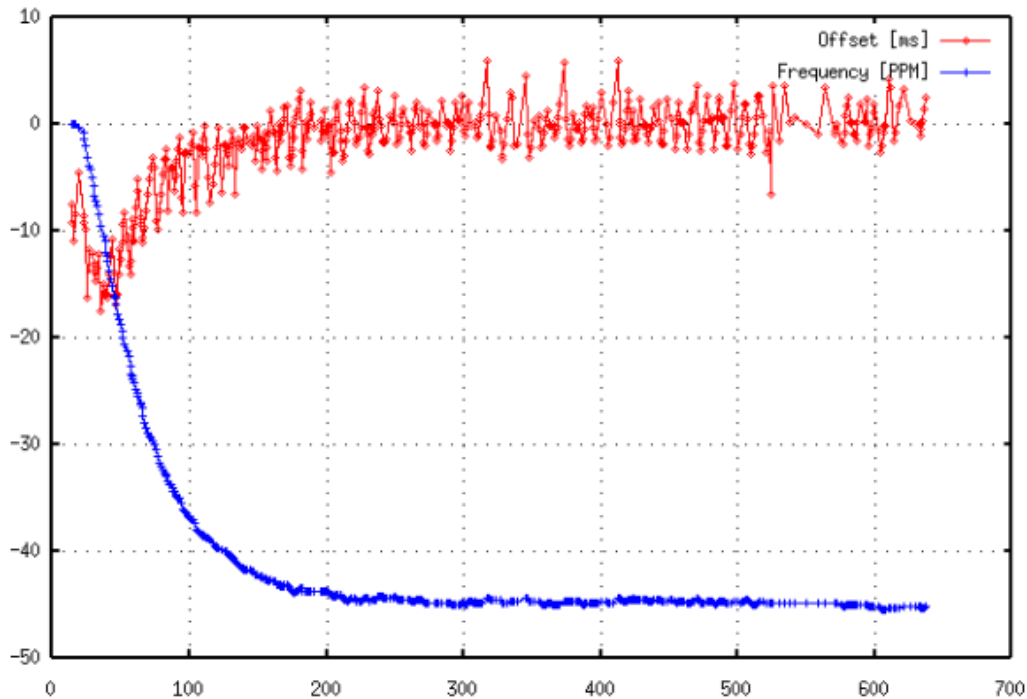

The above graph shows the offset recorded in the SNTP framework for three hour duration. The red liner in the graph shows the offset values retrieved through the ntpq query. Even though the clock accuracy seems good enough in the graph, we can see a higher deviation in between the offset entries. As per the calculations, SNTP has the clock accuracy within 100 milliseconds. All the calculation were based on the symmetrical consideration. That is there time taken from the client to server and server to client remains same. The offset values will be calculated based on the equations that has mentioned in the literature review section.

Now will have a look on the SPoT protocol offset representation. In this scenario, the experiment was conducted with two different IoT devices at the high noise level condition. The below graph illustrates the minimum, maximum and standard deviation of the offset representation of SNTP and SPoT protocols on both devices. The first graph represents the IoT device 1 and the second graph represents IoT device 2. The green liner in the graph represents the offset and can see almost a steady graph throughout the duration. This indicates that the standard deviation of offset errors is lower when compared to the SNTP protocol. The stability behind the standard deviation of offset is due to the rate synchronization implementation in SPoT.

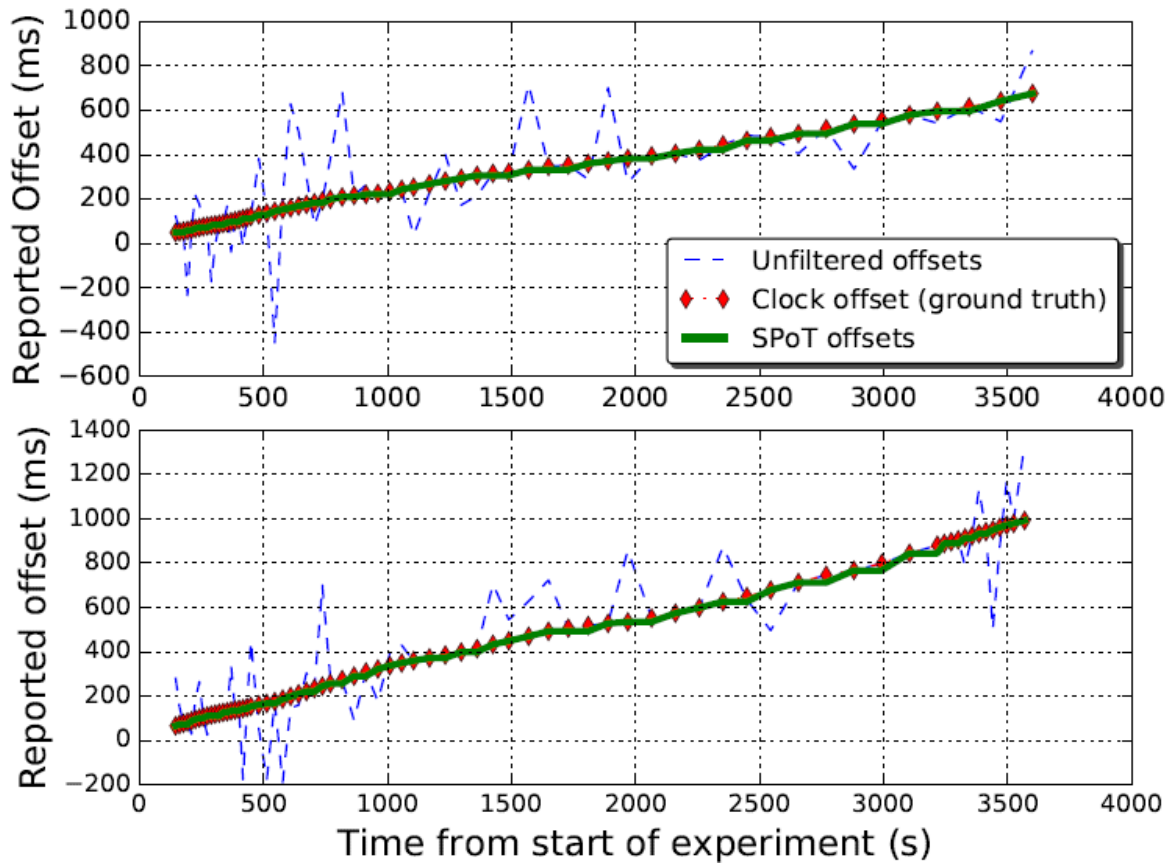

From the above graph, we can observe that, the minimum offset error statistic of both SNTP and SPoT protocols are zero milliseconds. Whereas the maximum offset value of the SNTP protocol reach up to 709 milliseconds and SPoT protocol reaches just 47.5 milliseconds. This difference itself shows the stability constraints of SNTP over SPoT protocol. While comparing the standard deviation of both protocols, we can find that SPoT is having around 7.9 milliseconds and the SNTP is having around 133.8 milliseconds offset errors. This shows SPoT is having lower standards deviation and 17 times more accurate than SNTP protocol at high noise levels. We can observe the same effects of offset calculations for both the IoT devices.

The research reveals the accuracy of both SNTP and SPoT protocol in terms of milliseconds at different noise levels. The figures that show the massive performance of SPoT protocol has been mentioned below.

| Protocol | Low Noise | Medium Noise | High Noise |
|----------|-----------|--------------|------------|
| SPoT     | 10.0      | 9.3          | 8.9        |
| SNTP     | 36.2      | 105.5        | 161.7      |

The time shown in the table is in milliseconds. We can notice that in each case, whether it is a low, medium or high, SPoT performed far better than the SNTP. It is clearly noted that SPoT protocol maintained its performance in all levels within the range approximate to 10 milliseconds. But SNTP has been increased drastically when compared the low noise level values to the medium and high noise level values.

The last question is about the parameter that determines the good time sync in the IoT platform. Basically, offset is the main parameter that determines the level of synchronization in both protocols. This value indicates the accuracy and reliability of the synchronization protocols. The findings of the third question evidence the reason and extend off the influence of offset in the time synchronization of both the protocols. Apart from these, the basic set of requirements or values for a good synchronization includes that the reference servers should have a proper sync based on the national standards. This can be achieved though the radio or atomic clock. The subnet that is been synchronized must be a reliable network. The synchronization protocol must be executing continuously to update the information in real-time.

After analysing all the potential solution that retrieved from the research, we came to the conclusion that SPoT performs 17 times more accurate and stable than SNTP protocol at various noise levels. That means, at different noise levels, SPoT maintains its clock accuracy of around 15 milliseconds whereas SNTP protocol fails to perform its stability at different levels. Figures show that SNTP performance is poor at high noise level when compared to low noise levels. The high noise levels of SNTP shows 161.7 milliseconds clock accuracy whereas for low noise levels it was just 36.2 milliseconds. This huge difference itself questions the stability and scalability of the SNTP protocols over the networks in IoT platform. At the same time, SPoT shows a clock accuracy of 10.0, 9.3 and 8.9 milliseconds accuracy for low, medium and high noise levels respectively. This shows the accurate, stable and scalable features of SPoT over SNTP protocol.

# Discussion

This research was a good initiative in the IoT platform for the time synchronization mechanism. It helps to have an in-depth knowledge on most of the concepts related to the time synchronization, especially in the SNTP and SPoT protocols. The key observation or outcome from this research is that the SNTP protocol have the clock accuracy of around 100 milliseconds, whereas the SPoT protocol has the accuracy range of approximately 15 milliseconds. In addition, the research proves that the SPoT performs 17x more accurate and stable than SNTP at different noise levels. These findings clearly sorting out the research question that arises at the beginning of this research study. This work starts with the basic analysis on the time synchronization in IoT and the different network protocols available for the time synchronizations. This review provides a list of network time protocols like SNTP, NTP, MQTT, PTP, and DTP that uses for various instances. The selection of SNTP and SPoT for the main research is due to the applicability and performance of both protocols in the IoT platform.

The first findings from this research study were about the synchronization method of SNTP and SPoT. As the previous section details, the nodes in the networks will adjust their internal clocks based on a reference server. For SNTP protocol, this will happened with the help of stratum 0 servers whereas, in SPoT, a SPoT server will act as a reference server. The Stratum 0 concept is widely used and accepted protocol for most of the IoT networks. This server will be in UTC time zone to maintain a unique and global standard for the network (Mills, 1991). The applicability of SNTP protocol is widely seen in the home networks to synchronize the computers used in. Most of the enterprise applications and IPTV application are using the SNTP protocol for the time synchronization. When compared the applicability of SPoT in the IoT platforms, we can observe that, environmental sensing is the most influencing field of IoT where SPoT system have a great future. It includes the temperature detection, smoke detection, indoor air quality monitoring and so on. Apart from this all SPoT also have a good future in the field of smart homes, smart grids, floor automation, smart-vehicles and gaming fields (Sathya Kumaran Mani, 2018).

The time calculation concept is really important to understand the time synchronization concept in IoT. Even though the calculations look similar for both the protocols, the algorithms used to execute or interpret these values are different. That is why we can see that SNTP only consider the symmetrical errors and SPoT considers both symmetric and asymmetric errors. SPoT algorithms also try to define the direction of the asymmetry by comparing the offset values. This calculation provides an effective and efficient time system throughout the network. In both protocols are analysed mainly based on the offset errors. This work closely related to the previous studies that examined the influence and effects of environmental factors like temperature and noise variations. An increased offset error results in a poor synchronization whereas a decreased offset error provides perfect sync to the network. Based on this theory, we come the outcome that Spot performs 17x more accurate than SNTP. Because SNTP has approximately 100ms offset value and not stable enough under various noise conditions. SPoT protocol always performs within the 10ms rage under different noise levels.

The main strength of this research study is to have a practical implementation of the SNTP protocol and analysis in different operating systems. The literature review for this protocol analysis provides a very good base for the practical implementation of SNTP. Because there has been many type pf research and works has been conducted to verify and evaluate the SNTP and NTP performances in the past years. This helped me a lot to go ahead with the SNTP experiences. Even though I had faced few issues to query the offset and delay details in Linux machines, the available works helped me to sort out this in an effective manner. SNTP is a widely accepted protocol for enterprise applications. That is the main strength behind this work. Because we were able to get all the required details easily. This proves the significance of the comparison of SNTP with the other protocols.

For every research studies, we will be having some limitations as well. When we start with this project, we were planned to have analysis on both SNTP and SPoT protocols and their implementations. But later on, due to some laboratory restrictions and the recent entry of SPoT protocol to the IoT platform became an obstacle for the practical implementation of SPoT over a network. Even though SPoT practically proves its time accuracy through the research experiments, the unavailability of more papers on SPoT made us in a dilemma to avoid the practical implementation. Instead, we planned to take the findings from the base research paper and analyse the values and finding thoroughly and compare those observations with the SNTP implementation results.

The Wireless Sensor Network (WNS) is one of the challenging and promising areas in recent years. Since many nodes are working together for a specific objective, data transfer and time synchronization are really important and it became one of the necessities of WSN. Because in the WSN platform, sensors will capture the data and transfer it to the microcontroller for further processing and convert into the readable format. Hence the time of the events that captured by each sensor plays an important role for each application that intended to deploy. Especially in environmental monitoring, video surveillance and asset tracking systems, the behaviour of the application are a real-time communication. Therefore timestamp of each node is really important and also to have a unique time throughout the network is more important in a system. This is really required to interpret the data collected effectively. The WSN network also possess the time parameters that we already discussed like clock drift, offset and time synchronisation. The Reference Broadcast Synchronization (RBS), Timing-sync Protocol for Sensor Network (TPSN) and Flooding Time Synchronization Protocol (FTSP) are the mainly used sync protocols in WSN networks (Ranganathan, P., & Nygard, K., 2010). Still, the researchers are looking forward for a strong algorithm to meet a secure and stable clock synchronization of the sensor network. Therefore, I believe this study and the performance of SPoT system concepts will enable the sensor networks to have an error-free time measurement system. For that, a study for the feasibility of SPoT protocol over the wireless sensor networks will help to evaluate the strengths and limitation of this applicability.

# Conclusion

In this paper, we found that SPoT protocol performs 17 times better than SNTP protocol in terms of clock accuracy over the network at different noise levels. Their accuracy difference has been calculated based on milliseconds. Since SPoT protocols considering both symmetric and asymmetric errors in its offset synchronization algorithm, it provides a more accurate time sync than SNTP protocol. The SPoT protocol not only identifies the asymmetric errors but also finds out the direction of asymmetry, magnitude of it and corrects the offset based on the direction and magnitude of asymmetry. But SNTP does not validate the asymmetric errors in the packet exchange and it considers all transactions under the symmetrical conditions. In addition, SPoT supports both thin and thick clients with varied computational capabilities and performance requirements. SNTP implementation and analysis result in the observation that it has the clock accuracy of approximately 100 milliseconds and will not be able to perform stable for different noise levels. Whereas this research shows that SPoT can perform stable at any noise levels and can maintain the clock accuracy within 15 milliseconds. All these observations help this study to meet its objective to find out the better protocol in IoT platform. This indicates that SPoT protocol is a promising development for the time synchronization issues in the IoT platform.

Internet of Things became a promising platform for these years. That is the reason why billions of devices are getting connected over the network for data transmission. Due to the increase in demand of the IoT platforms in the current society results in the emergence of many real-time applications including environmental monitoring, floor automation, smart home and gaming. The need of good time sync between the nodes for this application is the next level of thinking for many researchers for the successful operations of all these applications. This paper enables the readers to have a good learning on time synchronization in IoT concepts and the role of networking protocols in IoT to perform time synchronization. The increase in the number of IoT devices per year accelerates the motive to have a study on time sync between those devices. This proves the relevancy behind this research study at this point of time.

Extensive research on a system will help the system to grow up and outperform its best in the future. That was the motive behind this feasibility study of the SNTP and SPoT protocols. The feedback from the literature review for this study was based on the experiments and observations carried out as part of their work. In the same way, this work will also help the next level readers to have an idea about the time synchronization in IoT. In the future work, we planned to have the SPoT implementation in diverse environmental configuration with more IoT devices to evaluate the clock accuracy to hold a large number of nodes over the network. Furthermore, I strongly believe that this paper will help the future researchers to explore the time synchronization mechanism in the Wireless Sensor Network system with much easier. Because the WSN network also requires the proper time sync for their real-time data events captured by the sensors. In addition, this will give them a wide choice for the researchers for the implementation plan of various time synchronization protocols in WSN.

# References

- Borley, R. (2018). *What is The Internet of Things (IoT)?*. [online] Enterprise Software Development - Dootrix. Available at: https://dootrix.com/what-is-iot/ [Accessed 18 Oct. 2018].
- Burns, P. (2017). 5 Reasons Why Synchronization is Critical to IoT. Retrieved 05/08/2018, 2018, from iotforall, https://www.iotforall.com/iot-synchronization/
- Cox, D., Jovanov, E., & Milenkovic, A. (2005, March). Time synchronization for ZigBee networks. In *System Theory, 2005. SSST'05. Proceedings of the Thirty-Seventh Southeastern Symposium on* (pp. 135-138). IEEE.
- Doc.ntp.org. (2017). *ntpdate - set the date and time via NTP*. [online] Available at: http://doc.ntp.org/4.1.1/ntpdate.htm [Accessed 28 Oct. 2018].
- Elson, J., & Römer, K. (2003). Wireless sensor networks: A new regime for time synchronization. *ACM SIGCOMM Computer Communication Review*, *33*(1), 149-154.
- Elson, J., Girod, L., & Estrin, D. (2002). Fine-grained network time synchronization using reference broadcasts. *ACM SIGOPS Operating Systems Review*, *36*(SI), 147-163.
- Gartner. (2017). Gartner Says 8.4 Billion Connected "Things" Will Be in Use in 2017, Up 31 Percent From 2016. Retrieved 2018, from gartner.com, https://www.gartner.com/en/newsroom/press-releases/2017-02-07-gartner-says-8-billion-connected-things-will-be-in-use-in-2017-up-31-percent-from-2016
- Gayraud, R., & Lourdelet, B. (2010). *Network Time Protocol (NTP) Server Option for DHCPv6* (No. RFC 5908).
- GitHub. (2018). *jbenet/ios-ntp*. [online] Available at: https://github.com/jbenet/ios-ntp [Accessed 28 Oct. 2018].
- HU, J., & GAO, X. (2009). Application of SNTP-based time synchronization in digital substation. *Electric Power Automation Equipment*, *29*(3), 143-148.
- IoT Agenda. (2018). *What is internet of things (IoT)? - Definition from WhatIs.com*. [online] Available at:

- https://internetofthingsagenda.techtarget.com/definition/Internet-of-Things-IoT [Accessed 18 Oct. 2018].
- Jin, H., Zhang, M., & Tan, P. (2006, September). Clock synchronization integrated with traffic smoothing technique for distributed hard real-time systems. In *Computer and Information Technology, 2006. CIT'06. The Sixth IEEE International Conference on* (pp. 176-176). IEEE.
- Johannessen, S. (2004). Time synchronization in a local area network. *IEEE control systems*, *24*(2), 61-69.
- Khondokar Fida Hasan, C. W., Yanming Feng, Yu-Chu Tian. (2018). Time synchronization in vehicular ad-hoc networks: A syrvey on theory and practice. Retrieved from https://www.sciencedirect.com/science/article/pii/S2214209618300421
- Mani, S. K., Durairajan, R., Barford, P., & Sommers, J. (2018). A System for Clock Synchronization in an Internet of Things. *arXiv preprint arXiv:1806.02474*.
- Mills, D. (2005). *Simple network time protocol (SNTP) version 4 for IPv4, IPv6 and OSI* (No. RFC 4330).
- Mills, D. L. (1985). *Algorithms for synchronizing network clocks* (No. RFC 956).
- Mills, D. L. (1991). Internet time synchronization: the network time protocol. *IEEE Transactions on communications*, *39*(10), 1482-1493.
- Ntp.org. (2018). *How does it work?*. [online] Available at: http://www.ntp.org/ntpfaq/NTP-s-algo.htm#Q-REFCLK [Accessed 18 Oct. 2018].
- Park, J. K., & Kim, Y. T. (2008, September). An Enhanced SNTP (ESNTP) Clock Synchronization for High-Precision Network QoS Measurements. In *International Workshop on IP Operations and Management* (pp. 91-102). Springer, Berlin, Heidelberg.
- Ranganathan, P., & Nygard, K. (2010). Time synchronization in wireless sensor networks: a survey. *International journal of ubicomp*, *1*(2), 92-102.
- Rybaczyk, P. (2005). Network Time Protocol. Retrieved from https://link.springer.com/content/pdf/10.1007/978-1-4302-0039-0.pdf
- Sivrikaya, F., & Yener, B. (2004). Time synchronization in sensor networks: a survey. *IEEE network*, *18*(4), 45-50.

- Support.microsoft.com. (2018). [online] Available at: https://support.microsoft.com/en-au/help/262680/a-list-of-the-simple-network-time-protocol-sntp-time-servers-that-are [Accessed 28 Oct. 2018].
- Techopedia.com. (2018). *What is Simple Network Time Protocol (SNTP)? - Definition from Techopedia*. [online] Available at: https://www.techopedia.com/definition/4539/simple-network-time-protocol-sntp [Accessed 28 Oct. 2018].
- Tools.ietf.org. (1995). *RFC 958 - Network Time Protocol (NTP)*. [online] Available at: https://tools.ietf.org/html/rfc958 [Accessed 28 Oct. 2018].
- Van Greunen, J., & Rabaey, J. (2003, September). Lightweight time synchronization for sensor networks. In *Proceedings of the 2nd ACM international conference on Wireless sensor networks and applications* (pp. 11-19). ACM.
- Vasilakos, A. V., Spyropoulos, T., & Zhang, Y. (2016). *Delay tolerant networks: Protocols and applications*. CRC press.
- Waldron, D. (2014). SNTP or NTP ? Retrieved 06/08/2018, 2018, from galsys.co.uk, https://www.galsys.co.uk/news/sntp-vs-ntp/

# Appendix

## Appendix A – Reflection

The IFN702 unit really provides a fine set of learnings and practical experience for my future professional career. The experience was not only a unit oriented learning but it helped me to work in an organised and planned manner throughout the semester. The main advantage that I have earned through this unit is to progress according to the plan that we have created in Week 4. This helped to overcome the timeline slippages and analyses the progress of my work in each and every week.

- I believe that I have put forward my full effort for the successful SNTP implementation. When I started with the unit I was having a very little idea about the implementation part and it was interesting to learn things from Google and YouTube channels. I really hope that I have done the best in the SNTP implementation and analysis part along with the literature review by trying in both Windows and Linux platforms.
- Even though I have come up with the SPoT analysis I felt some incompleteness in the SPoT implementation part. Due to the laboratory restrictions, I was not able to recreate the scenarios for my research. I believe the experience will bring up the better results than from the literature reviews or literature observations. The unavailability of the requirement set insist me to take the literature observations for this part.
- The main challenge I faced was to trying the SNTP implementation in a Linux machine. Since the code and Linux properties were new to me, it was bit hard to implement and retrieve the results. But the YouTube videos and the learnings for IFN641 unit helped me to overcome this challenge in a greater extent. I am really happy to see the influence of other units for this project work.
- The overall experience was really a good learning for me. The experience with such protocol level was completely new to me. Therefore each and every step that I took was a learning for my professional development. This work really helped me to improve the problem-solving capabilities. Because while setting up SNTP in Linux, I was not getting the status as running initially due to the errors in the

- configuration file. I tried the different patterns to fix it up and it boosts me to gain a confidence in things.
- The weekly meetings with the supervisor helped me a lot to stick on the topic and guided me to follow the plan successfully. His supervision was really appreciable because he always enquired about the progress on my works to me on track.
- The area which I felt to improve is the ability to figure out the strong evidence for future developments. From this research study, I have contributed my findings and observations. But the future developments based on this study is written in very short since the lack of evidence on the claims.
- I really hope to work on a networking level in my future. There are many things around me to learn and many innovations are coming forward in every year in the networking field. It is not that easy to handle things in this level. The responsibilities are really risky and I wish to take that. This work enable me with courage to face the future steps in this field.
- Throughout the project, I made sure that I have attended all the weekly meetings with the supervisor and were updating the progress through the mails. My supervisor was always available through mail and I received the quick response for the queries I have raised.

Overall, the unit gives me the immense knowledge and a strong base for my future projects. It was really a great experience than the literature review that I have done as part of IFN701. I really felt the difference between the literature review and research from this work and were able to keenly observe many type of research as part of this work. It enhances my analytical skills and interpretation capabilities to a greater extent. IoT is really a vast topic. We will get a lot from different sources about IoT platform and related areas. From the vast topic, the ability to stick on a specific area and the learning the experience on that specified area is really appreciable. My supervisor helped me to select this topic by analysing my ability to do this work under his supervision. I really take this opportunity to thank my supervisor for his extreme support and coordination.

# Appendix B – Overall Project Plan

## GHANTT Chart

| GHANTT CHART | | | | | | | | | | | | | | | |
|---|---|---|---|---|---|---|---|---|---|---|---|---|---|---|---|
| **Task Details** | | | **Weeks** | | | | | | | | | | | | |
| Tasks | Duration in Weeks | Completion Status | 1 | 2 | 3 | 4 | 5 | 6 | 7 | 8 | 9 | 10 | 11 | 12 | 13 |
| Project Approval | 1 | 100% | ■ | | | | | | | | | | | | |
| Initial Meeting | 1 | 100% | | ■ | | | | | | | | | | | |
| Student Agreement | 12 | 100% | | ■ | ■ | ■ | ■ | ■ | ■ | ■ | ■ | ■ | ■ | ■ | ■ |
| Weekly Status Meeting | 13 | 100% | | ■ | ■ | | ■ | | | ■ | ■ | | ■ | ■ | |
| Initial Plan Presentation | 1 | 100% | | | ■ | | | | | | | | | | |
| Project Plan Submission | 1 | 100% | | | | ■ | | | | | | | | | |
| SNTP configuration setup | 3 | 100% | | | | | | ■ | ■ | | | | | | |
| SNTP protocol evaluation | 1 | 100% | | | | | | | ■ | | | | | | |
| Experimental observation | 1 | 100% | | | | | | | | ■ | | | | | |
| SPoT Configuration Details | 2 | 100% | | | | | | | | | ■ | | | | |
| SPoT Implementation Details | 1 | 100% | | | | | | | | | ■ | | | | |
| Compare the SNTP and SPoT | 3 | 100% | | | | | | | | | ■ | ■ | | | |
| Summarise all findings | 3 | 100% | | | | | | | | | | | ■ | ■ | ■ |
| Final Presentation | 1 | 100% | | | | | | | | | | | | ■ | |
| Project Report Creation | 7 | 100% | | | | | | | | | ■ | ■ | ■ | ■ | ■ |
| Project Report Submission | 1 | 100% | | | | | | | | | | | | ■ | |

# Appendix C – Project Log Sheets
## Log Sheet 1

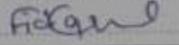

## Log Sheet 2

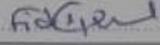

## Log Sheet 3

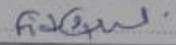

## Log Sheet 4

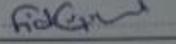

**Log Sheet 5**

Student's Name: Nelda Raju       Date: 08/10/2018       Meeting No: 5

Project title: Feasibility Study on SNTP and SPoT Protocol on IoT Platform       UNIT: IFN702

Supervisor's Name: Fida Hassan       Supervisor's Signature: [signed]

Journal entry logged into Blackboard (Optional)

Update on progress since last meeting, and challenges faced if any (noted by student before mandatory supervisory meeting):
1. Comparison result of both SNTP and SPoT

Items for discussion (noted by student before mandatory supervisory meeting):
1. Discuss about the findings and observations.
2. Discuss on the project presentation

Action List (to be attempted or completed by student by the next mandatory supervisory meeting):
1. Create more visual aids to prove the findings

**Log Sheet 6**

Student's Name: Nelda Raju       Date: 18/10/2018       Meeting No: 6

Project title: Feasibility Study on SNTP and SPoT Protocol on IoT Platform       UNIT: IFN702

Supervisor's Name: Fida Hassan       Supervisor's Signature: [signed]

Journal entry logged into Blackboard (Optional)

Update on progress since last meeting, and challenges faced if any (noted by student before mandatory supervisory meeting):
1. Ready with the presentation slides

Items for discussion (noted by student before mandatory supervisory meeting):
1. Discuss about the presentation topics.
2. Discuss on related work details

Action List (to be attempted or completed by student by the next mandatory supervisory meeting):
1. Draft the project report
2. Include all the findings and observations

**Log Sheet 7**

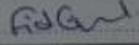